\documentclass[11pt,letterpaper]{article}
\usepackage[top=0.85in,left=0.75in,footskip=0.75in]{geometry}
\usepackage[utf8]{inputenc}
\usepackage{authblk}
\usepackage[sorting=none]{biblatex}
\usepackage{graphicx}
\usepackage{amsmath}
\usepackage{lineno}
\usepackage{color}

\title{Interaction of functional brain networks is formed by $k$-clique percolation in the human structural connectome}

\author[1,2]{Vasilii Tiselko}
\author[3*]{Olesia Dogonasheva}
\author[1]{Artem Myshkin}
\author[1,4]{Olga Valba}

\affil[1]{Laboratory of Complex Networks, Center for Neurophysics and Neuromorphic Technologies, Moscow, Russia}
\affil[2]{Centre for Cognition and Decision making, Institute for Cognitive Neuroscience, National Research University HSE, Moscow, Russia}
\affil[3]{Université Paris Cité, Institut Pasteur, AP-HP, Inserm, Fondation Pour l'Audition, Institut de l’Audition, IHU reConnect, F-75012 Paris, France}
\affil[4]{National Research University HSE, Moscow, Russia}
\affil[ ]{ }
\affil[*] {olesia.dogonasheva@pasteur.fr}
\date{}                     
\makeatletter
\renewcommand\@biblabel[1]{#1.}
\makeatother

\begin{document}
\maketitle

\begin{abstract}

The human structural connectome has a complex internal community organization, characterized by a high degree of overlap and related to functional and cognitive phenomena. We explored connectivity properties in connectome networks and showed that $k$-clique percolation of an anomalously high order is characteristic of the human structural connectome. The resulting structural organization maintains a high local density of connectivity distributed throughout the connectome while preserving the overall sparsity of the network. To analyze these findings, we proposed a novel model for the emergence of high-order clique percolation during network formation with a phase transition dynamic under constraints on connection length. Investigating the structural basis of functional brain subnetworks, we identified a direct relationship between their interaction and the formation of clique clusters within their structural connections. Based on these findings, we hypothesize that the percolating clique cluster serves as a distributed boundary between interacting functional subnetworks, showing the complex, complementary nature of their structural connections. We also examined the difference between individual-specific and common structural connections and found that the latter plays a sustaining role in the connectivity of structural communities. At the same time, the superiority of individual connections, in contrast to common ones, creates variability in the interaction of functional brain subnetworks.

\end{abstract}

\section{Introduction}

The anatomical and functional connection networks of the human brain are increasingly studied through the lens of network science, based on graph theory techniques and terminology \cite{guerra2021homological, szalkai2015budapest, Zeeman1965TopologyOT}. This approach provides a powerful framework for understanding the complex structural organization of brain connections.

In this network representation, a structural connectome depicts the brain's anatomical connection network. Small regions of gray matter are referred to as nodes, while edges represent the axon fibers connecting these regions, as identified through diffusion MRI \cite{tzourio2002automated,fan2016human}. This model allows for a comprehensive analysis of brain connectivity patterns.

Complex networks, including brain networks, often exhibit meso-scale or global structural features. Of particular interest is the community structure, characterized by densely connected node communities with sparse or weak inter-community connections \cite{girvan2002,fortunato2010}. In the structural connectome, the intersection and interaction of these dense clusters are believed to underpin various cognitive phenomena \cite{bullmore2012, sporns2016}. However, the organizational principles governing the community architecture of human brain networks remain poorly understood.

Recent studies have shed light on some aspects of this architecture. For instance, study \cite{lee2022} demonstrated that structural connectomes exhibit a high degree of community overlap, which correlates with cognitive flexibility. This analysis employs both the Louvain community detection algorithm and the link community algorithm to uncover community overlaps in the brain. Thus, while these studies have shed some light on the brain's network architecture, they often fail to explain the underlying mechanisms that lead to the formation and interaction of these communities, particularly at higher levels of connectivity.

An alternative approach to analyzing overlapping community structures is the clique percolation method \cite{derenyi2005clique,palla2007critical}. This method, which examines the percolation of $k$-cliques, offers a more nuanced view of network organization. Unlike traditional community detection algorithms that often assign nodes to a single community, the clique percolation method allows for the identification of overlapping communities. This is particularly advantageous in brain networks, where regions may participate in multiple functional or structural communities simultaneously. While the critical link probability for $k$-clique percolation can be analytically determined for random Erdős-Rényi networks, estimating this threshold for real-world networks, such as brain connectomes, presents a significant challenge.

The concept of $k$-clique percolation has recently gained traction in cognitive and semantic network analysis. Studies \cite{cosgrove2021, valba2022} have applied this method to investigate aging in semantic memory and the capacity limits of working memory, respectively. These applications demonstrate the method's potential for unraveling complex cognitive processes, particularly in cases where brain regions may contribute to multiple cognitive functions.

In this paper, we analyze the community organization of human connectomes, focusing on the dataset described in \cite{kerepesi2017braingraph}. This dataset comprises structural connectomes of $426$ human subjects, derived from the Human Connectome Project. The connectomes are available in five resolutions, representing anatomically identified regions of interest in the brain. 

Also, in this study, we propose a novel rule for network evolution, which can be viewed as a modification of exponential random graphs under metric constraints. This mechanism yields clique communities of orders consistent with observed data. Our approach builds upon existing models of brain network development, such as the trade-off between wiring cost and topological complexity \cite{bullmore2012}. However, our model introduces additional constraints on connection length, reflecting the biological limitations of axonal projections in the brain. This extension allows us to more accurately capture the formation of high-order clique structures observed in human structural connectomes while still maintaining the principles of efficient information processing and metabolic cost minimization that are fundamental to brain organization.

To analyze how structural organization of the human brain reflects functionality, we used data on functional subnetworks \cite{yeo2011organization}. Our approach to analyzing the organization of structural communities reveals a complex interplay between observed structural features and the interaction of functional subnetworks in the human brain. By investigating the relationship between these functional subnetworks and the structural features we observe, we aim to bridge the gap between structural and functional connectivity, potentially offering new insights into cognitive processes and brain disorders. For instance, recent studies have shown that alterations in structural connectivity patterns are associated with various neurological and psychiatric conditions. The study \cite{fornito2015connectomics} demonstrated that schizophrenia patients exhibit disruptions in the rich club organization of brain networks, which may contribute to cognitive deficits. Similarly, it was found that brain disorders target specific network hubs \cite{crossley2014hubs}, suggesting that our understanding of structural-functional relationships could inform targeted interventions. By analyzing the relationship between structural connectivity and functional subnetworks, our research may contribute to the development of more precise diagnostic tools and personalized treatment strategies for a range of brain disorders.

\section{Methods}

\subsection{Data description} 
\paragraph{Structural connectome data.}
The Human Connectome Project (HCP) has provided high-resolution structural connectome data from individuals aged $22$ to $35$ years \cite{mcnab2013human, kerepesi2017braingraph}. This dataset comprises structural connectomes of $426$ human subjects. The connectomes are available in five resolutions: $83$, $129$, $234$, $463$, and $1015$ nodes, each representing anatomically identified regions of interest (ROIs) in the brain. These multi-resolution graphs provide a view of brain connectivity, with nodes representing specific anatomical areas and edges depicting the neural fibers connecting them.

Subcortical structures have widespread connections across the brain, often appearing as hubs in network representations. In our analysis, we identified a densely connected subnetwork comprising vertices that represent hubs and belong to subcortical structures. To focus on the broader network architecture, this densely connected subnetwork was excluded from further analyses across the entire ensemble of networks.

\paragraph{Functional subnetworks data.}
To analyze the relationship between structural connectivity and functional organization we used functional subnetworks \cite{yeo2011organization}. This dataset comprises functional magnetic resonance imaging (fMRI) data from a large cohort of $1000$ healthy young adults of age $18$-$35$ years. The fMRI data were aligned to a common surface-based coordinate system, preserving the cortical surface topology. The resulting dataset provides a parcellation of the human cerebral cortex based on intrinsic functional connectivity. We mapped the vertices of our structural connectomes onto these functional subnetworks using the same parcellation atlas \cite{lausanne}, creating a bijection between structural connectome vertices and functional subnetworks. In this partition, each node of the functional connectome is assigned to exactly one functional subnetwork \cite{yeo2011organization}. 

\subsection{$K$-clique percolation} 

A \emph{k-clique} is defined as a complete subgraph consisting of $k$ vertices, where every vertex is connected to every other vertex within the subgraph \cite{derenyi2005clique, palla2007critical}. Two $k$-cliques are considered \emph{adjacent} if they share exactly $k-1$ vertices, differing only by a single vertex. A \emph{k-clique chain} is formed by a sequence of pairwise adjacent $k$-cliques. When two $k$-cliques can be connected by at least one $k$-clique chain, which represents a sequence of $k$-cliques where each adjacent pair shares $k-1$ vertices, they are called \emph{k-clique-connected}. Finally, a \emph{k-clique percolation cluster} represents the maximal $k$-clique-connected subgraph, encompassing all $k$-cliques that are $k$-clique-connected to a specific $k$-clique (Fig.\ref{fig1}a).

The concept of $k$-clique percolation is particularly interesting when applied to random graph models, such as Erdős-Rényi graphs. These graphs exhibit a series of phase transitions as the probability $p$ of connection between any two nodes increases. For $k=2$, this transition is well understood and manifests as the emergence of a giant component when the critical probability $p_c(k=2)=\frac{1}{N}$ is reached, where $N$ denotes the number of nodes in the network. This phenomenon generalizes to higher values of $k$, where for each $k$, there exists a threshold probability $p_c(k)$ above which $k$-cliques organize into a giant community \cite{palla2007critical}:

$$
p_c(k)=\frac{1}{\left[ N(k-1)\right]^{1/(k-1)}}.
$$

The study of $k$-clique percolation in random graphs provides a useful benchmark for understanding the organization of dense substructures in more complex networks, such as brain connectomes. By comparing the percolation properties of empirical networks to those of random graphs, we can identify non-random features that may be functionally significant. For instance, the presence of high-order clique percolation in brain networks at densities where such structures would be unlikely in random graphs can indicate the presence of functionally important modules or processing units \cite{bullmore2009complex}.

\subsection{The edge confidence}

To quantify the reliability of connections across subjects, we used an edge confidence measure, similar to the methodology proposed in \cite{szalkai2015budapest, szalkai2017parameterizable}. The weight of an edge was calculated as its probability of occurrence across the entire set of networks. This approach helps to distinguish between common connections in the brain network and more individual-specific ones. The resulting edge distribution with respect to confidence values is broad and uniform, with a characteristic peak for the weakest connections and a decline for the most common connections in the population (Fig.~\ref{fig1}b).

\subsection{Network decomposition and inverse decomposition}

To investigate how structural properties depend on edge confidence, we used two complementary approaches: network \textit{decomposition} and \textit{inverse decomposition}. In network decomposition, we applied a cutoff threshold $\tau$ for edge confidence, removing all links with weights below this threshold. Conversely, in inverse decomposition, we removed connections with weights greater than or equal to a threshold $\theta$, allowing us to analyze subgraphs of weak (individual) links. The network decomposition and inverse decomposition techniques allow analyzing how different structural properties shape the organization of the connectome.

\subsection{Network model} 

The model can be viewed as a modification of exponential random graphs under metric constraints. The model starts as a random network embedded within a 3-dimensional sphere, providing a spatial framework where connection lengths are defined as the Euclidean distance between nodes. Each initial network is randomly selected from the set of all graphs containing $100$ nodes and $1000$ edges, and nodes are evenly distributed inside a sphere of unit radius. The network structure changes dynamically through edge rewiring, controlled by chemical potential and constrained by limits on the allowable connection lengths (maximal connection length). The evolution of the network structure optimizes the growth of recurrent connections, saturating the network with triangles while maintaining its edge density. At each evolutionary step, a new network configuration with a rewired connection is accepted based on either the condition of an increase in network transitivity or a probabilistic condition similar to the Metropolis optimization approach, depending on the chemical potential $\mu$ (the value of the exponent of the product of the chemical potential and the difference between the transitivity values of the two configurations is greater than a random, uniformly selected number from the interval $[0, 1]$). Transitivity is calculated as the ratio of the number of triangles to the number of triads (two edges with a single shared vertex). The evolution continues until a fixed transitivity value is reached or the computational complexity is exponentially increased for further transitivity increments. In our simulations, we typically limited the maximum transitivity to $0.6$, slightly higher than the average of $0.5$ observed in human connectomes.

\subsection{Network embedding} 

We used an approach of network embedding into metric space, in particular, a hyperbolic embedding. It has recently been shown that human connectomes can be anatomically analyzed using hyperbolic embedding \cite{allard2020navigable, whi2022hyperbolic, zheng2020geometric, tadic2019functional}. In the embedding process, we determine a set of hyperbolic coordinates for each node. One of the coordinates is associated with the number of node connections, and the other angular-wise coordinate is related to the degree of similarity based solely on the structural organization of connections. It is the internal hierarchy of the structure that determines the naturalness of the embedding of this type, which is described in detail in \cite{SerranokrioukovBoguna2010hiddenMetSp, krioukov2010hypcomnet, zheng2020geometric}. We considered the embedding of networks into a minimal two-dimensional hyperbolic space using the method described in \cite{hyp_embed2017}, thus obtaining one of the coordinates reflecting the node's similarity. When embedding networks, we cut off all connections with edge confidence less than $0.2$ in each connectome to ensure that the common structure between all connectomes had a greater influence during the embedding process.

\section{Results}

We study the internal connectivity of human structural connectomes and its relation to functional subnetworks, presenting our findings as follows. We start by showing that high-order $k$-clique percolation is a distinctive feature of the human structural connectome. To understand the observed phenomenon, we proposed a novel model for emerging high-order $k$-clique percolation with phase-transition dynamics under specific constraints on connection length. We then examined the structural connections underlying known functional subnetworks of the brain, showing how the interaction between functional subnetworks related to the formation of structural clique communities. In the last section, we explored the differences between individual and common structural connections, employing network \emph{decomposition} and \emph{inverse decomposition} techniques (see the Methods section for details).

\begin{figure}[!ht]
\centerline{\includegraphics[width=1\linewidth]{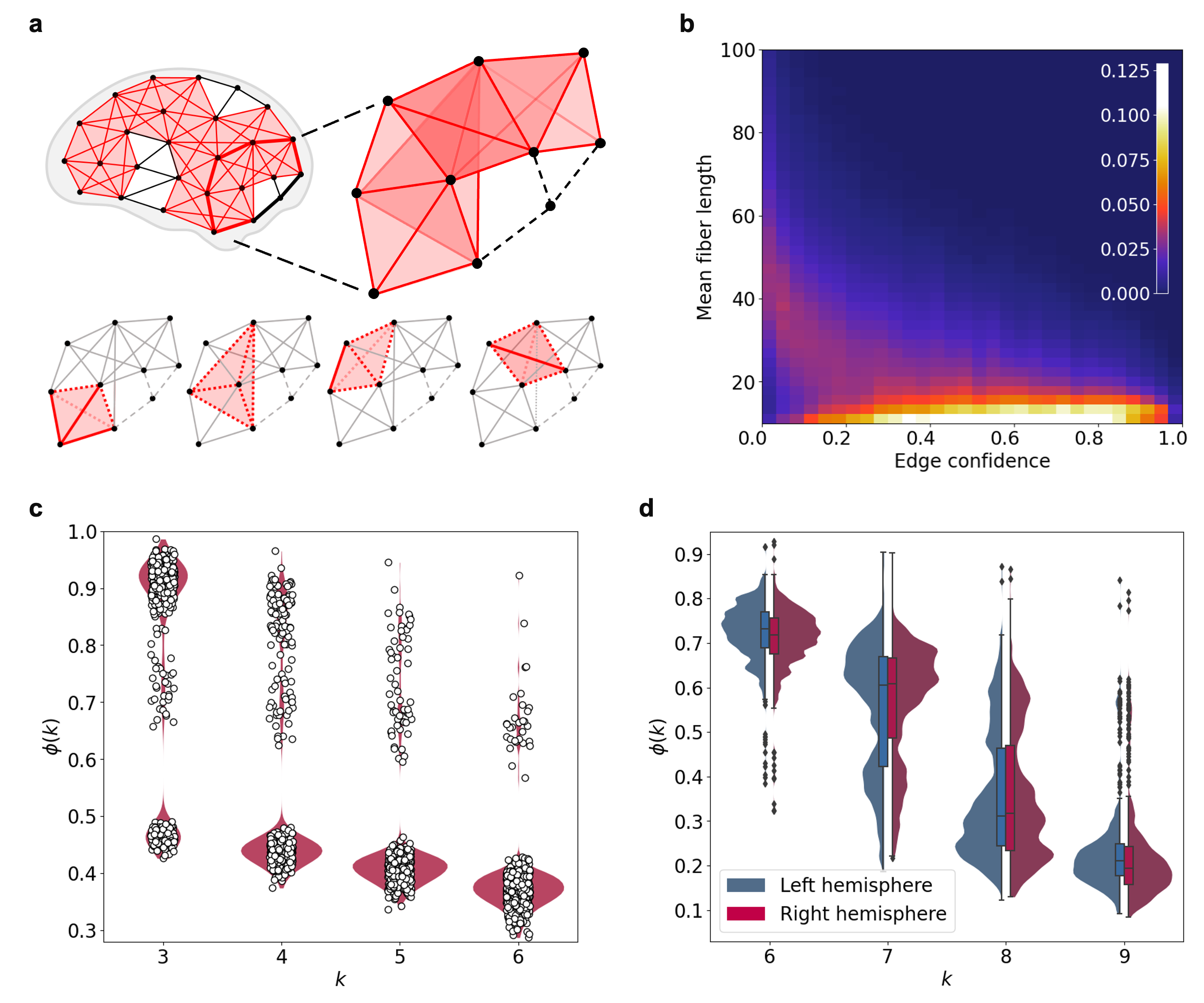}}
\caption{
(a) The structure of the percolation $k$-clique cluster occupies the entire volume of the network and consists of merged $k$-cliques. (b) The distribution of structural connectome connections by the length of the forming fibers and edge confidence. (c) The share of the $k$-clique cluster $\phi(k)$ from the entire connectome network for different clique orders $k$ (dots represent individual structural connectomes). (d) The share of the clique cluster $\phi(k)$ in brain hemispheres.}
\label{fig1}
\end{figure}

\subsection{Community structure and $k$-clique percolation in human structural connectomes.} 

\paragraph{Higher-order clique percolation in hemispheric networks.}

Investigating $k$-clique clusters of different orders $k$, we examine the connectivity of dense structures and their internal organization, even when these structures are poorly separable. The emergence of large-scale structures consisting of merged $k$-cliques of high order $k$ (i.e., cliques on $k$ vertices) indicates a more complex structural organization or a high connection density. In various scientific fields, the formation of such structures is often associated with the emergence of a new system state \cite{palla2007critical}.

The emergence of clique percolation in known models depends primarily on the connection density, which is especially interesting in random network models. The observed hemispheric connection density in connectomes (the proportion of actual connections compared to the total possible connections) satisfies $p_c(k = 3) < \rho < p_c(k = 4)$. It suggests that if connections within the hemispheric network were formed randomly, following an Erdős-Rényi-like model, we would expect most vertices to be included in the percolation cluster at $k=3$, with higher-order percolation clique clusters being absent. In addition, such models consider the formation of long-range connections, while in connectomes, on the contrary, we observe that the vast majority of connections are short (Fig.\ref{fig1}b, \emph{edge confidence} shows the frequency of connection occurrence in the network sample). 

In connectome networks, we observe the percolation of $k$-cliques of fairly high orders, up to $k=7$ in individual hemispheres (Fig.~\ref{fig1}d). The order of the percolating cluster $k$ in the whole connectome is limited by interhemispheric connections, apparently not forming clique structures of high orders. The observed order of $k$-clique percolation in the hemispheres is phenomenal, given the density of connections, and clearly shows the inconsistency of simple random models to describe intracerebral connectivity (Fig.~\ref{fig1}d).

\paragraph{A novel network model for high-order $k$-clique percolation in the human structural connectome.}

As observed, the human structural connectome exhibits a predominance of short connections, particularly for those with high edge confidence (Fig.~\ref{fig1}b). Despite this characteristic, the overall connection density appears insufficient for the emergence of high-order $k$-clique clusters typically observed in random networks. To reconcile these observations, we developed a novel network model that incorporates metric constraints on the length of the connection during structural formation. In the model, the network is formed dynamically, where the growth of recurrent connections is stimulated in the network structure, saturating the network with triangles while maintaining its edge density (Fig.\ref{fig3}c). The structure changes probabilistically due to rewiring of connections, controlled by the chemical potential parameter and restrictions on the length of new connections (see details in Methods).

\begin{figure}[ht!]
\centerline{\includegraphics[width=1\linewidth]{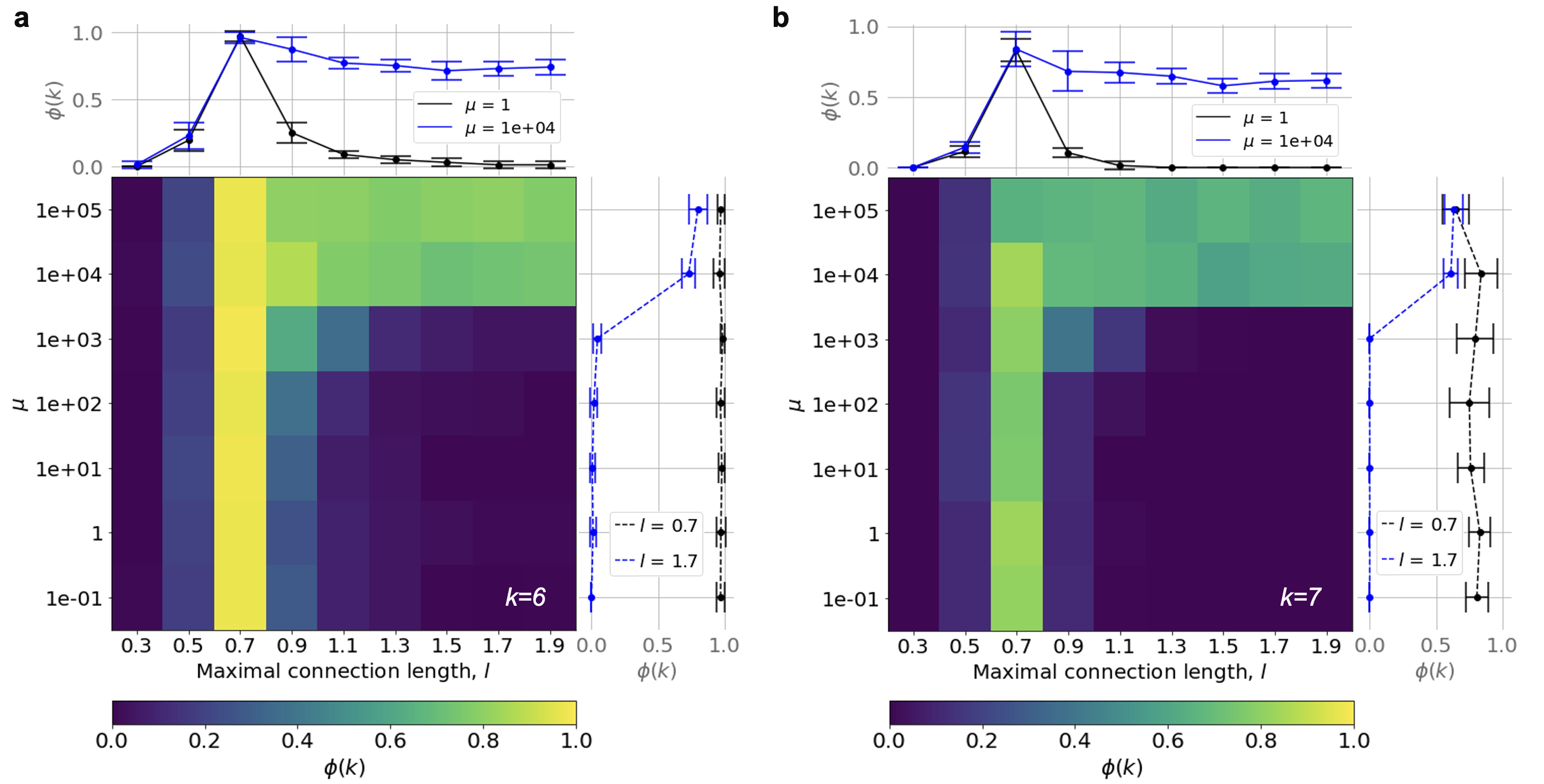}}
\caption{$k$-clique cluster share of the model networks $\phi(k)$ for clique orders $k=6$ (a) and $k=7$ (b), depending on model parameters (each value is averaged over $10$ networks). The networks were generated through an evolutionary process that optimizes the growth of recurrent connections under constraints on permissible connection length (\emph{maximal connection length}) and a fixed value of the chemical potential $\mu$. Large-order $k$-clique percolation, characteristic of human structural connectomes, occurs in a certain parametric domain. The side graphs show $\phi(k)$ change for fixed values of one of the parameters.}
\label{fig2}
\end{figure}

Figure~\ref{fig2} shows the maximal network share $\phi(k)$ occupied by the $k$-clique cluster for $k= 6,7$, as a function of the chemical potential and the factor limiting permissible connection length (\emph{maximal connection length}). The density of connections in the model networks, as for connectomes, suggests the formation of $k$-clique clusters of maximum order of only $k=3$ or $k=4$. Remarkably, in a certain range of model parameters, we observe the persistent emergence of $k$-cliques percolation of order up to $k=7$ (Fig.~\ref{fig2}, maximal edge length $\approx 0.7$; Fig.~\ref{fig3}b). The fraction occupied by the clique cluster $\phi(k)$ is usually considered as an order parameter for percolation. A structural phase transition in a network occurs when the permissible connection length is limited to a certain range, which is quantitatively dependent on network density (Fig.~\ref{fig2}, parametric domain where cluster size is close to $\phi(k)=1$). 

\begin{figure}[ht!]
\centerline{\includegraphics[width=0.8\linewidth]{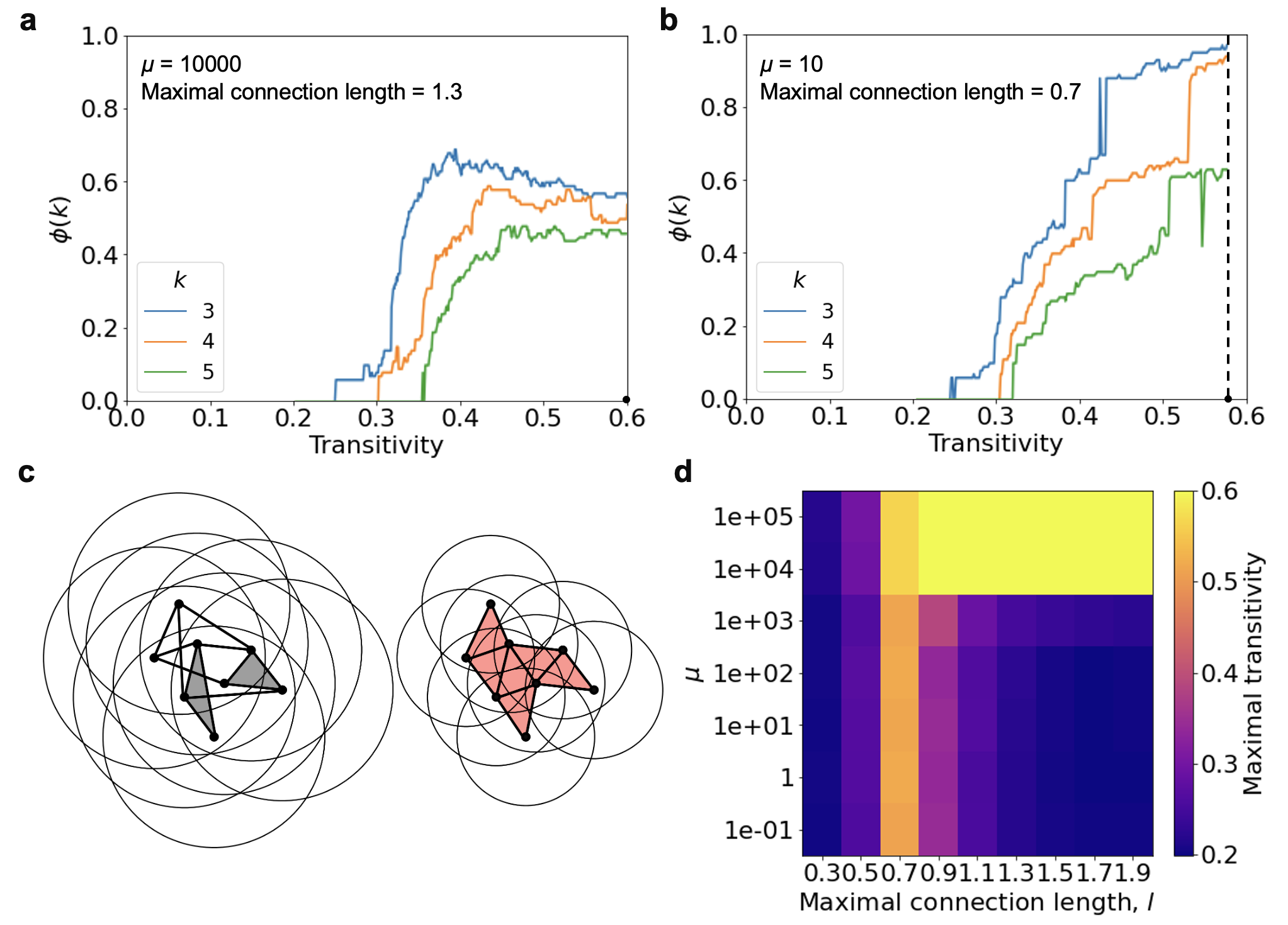}}
\caption{
(a-b) Characteristic dynamics of the $k$-clique clusters fraction $\phi(k)$ during model network formation with increasing transitivity. (a) At high chemical potentials, where rewired connections predominantly increase transitivity, the network tends to form dense clusters that are not necessarily interconnected. (b) When the length of connections is limited to a certain range, a percolation cluster is formed even at low chemical potential. (c) Schematic illustration of the change in the network structure during random edge rewiring with different permissible connection lengths. (d) The maximal value of transitivity when forming a network depends on the model parameters (each value is averaged over $10$ networks).
}
\label{fig3}
\end{figure}

In our model, low chemical potential corresponds to a high probability of edge rewiring regardless of the contribution to transitivity, i.e., the accepting of non-optimal states (contrary to typical optimization problems due to the sign of the transitivity difference). Under these conditions, edge rewiring occurs almost indiscriminately, regardless of changes in transitivity, resulting in the generation of a new random graph on a fixed set of vertices at nearly every step. Consequently, we can analytically estimate the expected maximal order of the percolation cluster $\rho_c$ for such random graphs. Transitivity in this case cannot increase significantly (Fig.~\ref{fig3}a,b). However, when the connection length is limited to a certain range, percolation still occurs, which is shown schematically in Figure~\ref{fig3}d.

At high chemical potentials, where rewired connections predominantly increase transitivity, the network tends to form dense clusters that are not necessarily interconnected. The global optimum of such dynamics is a clustered network, where clusters occupy only a fraction of the network volume and are tightly packed (Fig.~\ref{fig3}a). As a result, $k$-cliques do not percolate for large $k$, despite network transitivity reaching its maximal values (Fig.~\ref{fig3}a). However, under certain constraints on the connection length, clique percolation occurs as before. In Figure~\ref{fig3}b, showing the maximum achievable transitivity of the network, it is observed that in the percolation range within connection length limits, the transitivity is slightly lower than when forming clusters without constraints (Fig.~\ref{fig3}d). Thus, metric constraints on the connection length under such dynamics locally limit the maximum connection density.

\subsection{Clique community organization reveals an interaction between functional brain subnetworks at the level of structural connections.}

\paragraph{High structural intertwining of functional subnetworks.}

Recent research revealed several functionally connected subnetworks within human functional connectomes, characterized by a high degree of separability (i.e., reproducibly distinguishable based on their connectivity patterns)\cite{yeo2011organization}. We studied the structural connections underlying these functionally separated subnetworks using the same anatomical parcellation \cite{cammoun2012mapping}.

\begin{figure}[ht!]
\centerline{\includegraphics[width=1\linewidth]{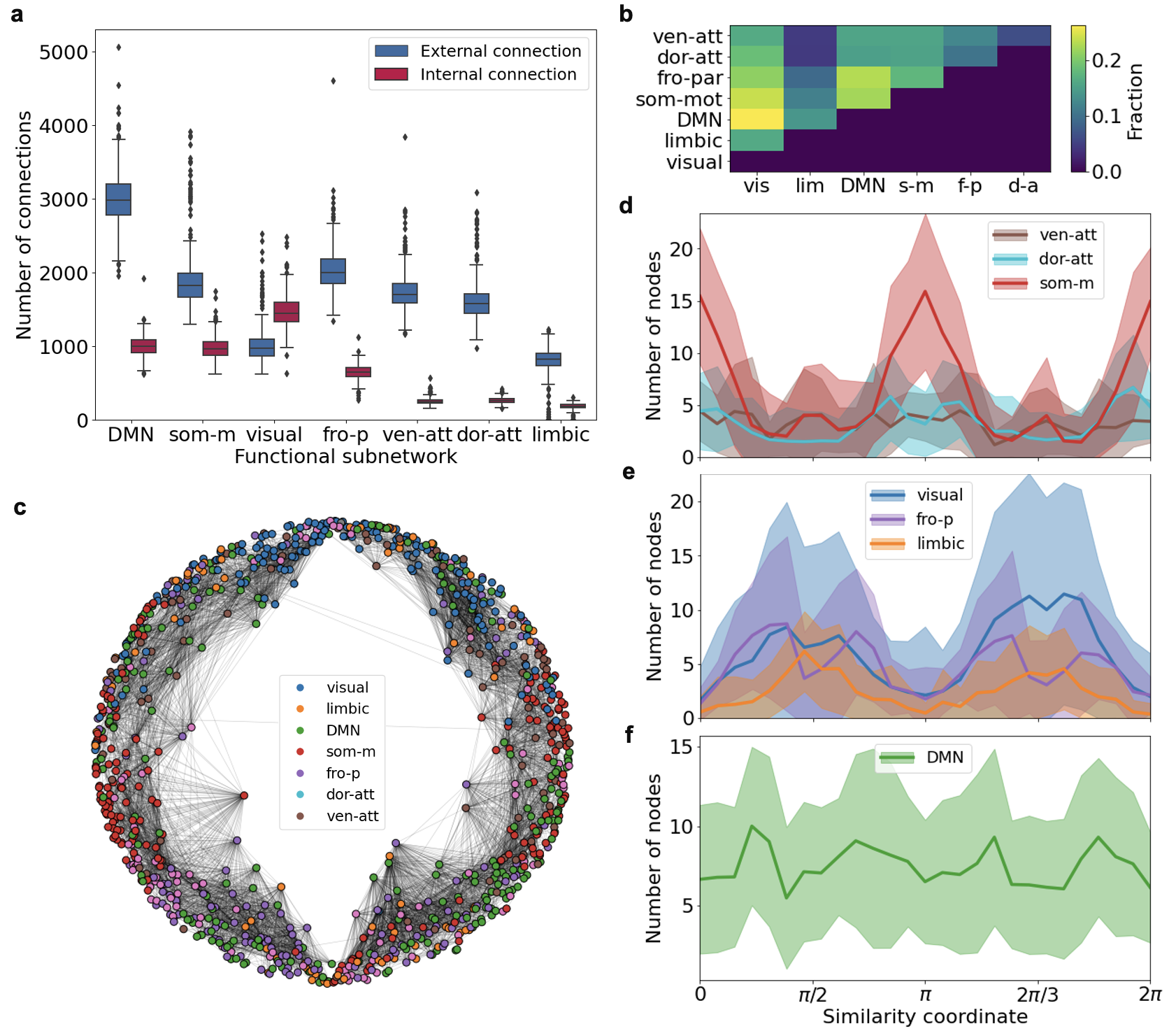}}
\caption{ 
(a) The number of external and internal structural connections of functional subnetworks of the human brain. (b) The fraction of external connections connecting functional subnetworks in pairs (each value is averaged over the entire sample of connectomes). (c) Hyperbolic embedding of the structural connectome allows studying the similarity between nodes based solely on the structural properties of the network. Colors of nodes correspond to different functional subnetworks. (d-f) Distribution of connectome nodes along similarity coordinate for different functional subnetworks: \emph{ventral-attention, dorsal-attention, somato-motor} (d), \emph{visual, fronto-parietal, limbic} (e), \emph{default mode network (DMN)} (f). 
}
\label{fig4}
\end{figure}

We distinguished between internal and external subnetwork connections. Internal connections are edges in the structural connectome where both vertices belong to the same functional subnetwork. External connections, on the other hand, are edges where only one vertex belongs to a functional subnetwork. Notably, some external connections become internal when considering the union of two or more subnetworks. For all subnetworks, except the \emph{visual} network, external connections significantly outnumber internal ones (Fig.~\ref{fig4}a). This observation suggests that functional subnetworks are not well-defined structural clusters but rather interact extensively with each other.

The proportion of edges that connect networks in pairs from all their external edges is small and similar for all subnetworks, which also shows dense mutual connectivity between all subnetworks (Fig.~\ref{fig4}b). Given such dense interaction of subnetworks, we explored the possibility of observing their functional proximity based solely on the structural organization. 

It has recently been shown that human connectomes can be anatomically analyzed using hyperbolic embedding \cite{allard2020navigable, whi2022hyperbolic, zheng2020geometric, tadic2019functional}. We use a similar approach and obtain node similarity from their angular-wise coordinate in the embedding, based solely on the internal structural properties of the network (see the methods section for details). Thus, we want to explore the relationship between the functional affiliation of nodes and the measure of their similarity from the geometric properties of the network structure (Fig.~\ref{fig4}c). Figure~\ref{fig4}c shows the embedding for one of the connectomes. Each node acquires two coordinates in a two-dimensional hyperbolic embedding, where the radial coordinate is related to the node hubness and the second angular coordinate can be used as a measure of node similarity, determined by the structural organization of the connections. The node's colors correspond to the functional subnetworks to which they belong, and the clustering of nodes by color along the boundary of the embedding is visually distinguishable.

Observing the distribution of nodes of each functional subnetwork by the similarity coordinate, we observe the complexly mutual complementary patterns (Fig.~\ref{fig4}d-f). Excepting the \emph{DMN}, two groups can be roughly distinguished among all functional subnetworks by the similarity of the distribution pattern (Fig.~\ref{fig4}d,e). In each group, the distribution pattern is periodic, clearly reflecting the hemispheres' separation. Interestingly, within each hemisphere, the structure is reproduced within the sample of networks and, in a sense, complements each other due to the coincidence of density bumps according to the similarity. The \emph{DMN} network in turn is distributed more evenly, merged with all other subnetworks at once, and ambiguously relates to each of the previously identified groups (Fig.~\ref{fig4}f). This finding aligns with previous research emphasizing how the \emph{DMN}'s support information integration across various brain regions \cite{raichle2015restless}. As a result, we observe the grouping of known functional subnetworks by a completely different method, based solely on the internal structure of connections.

\begin{figure}[ht!]
\centerline{\includegraphics[width=0.8\linewidth]{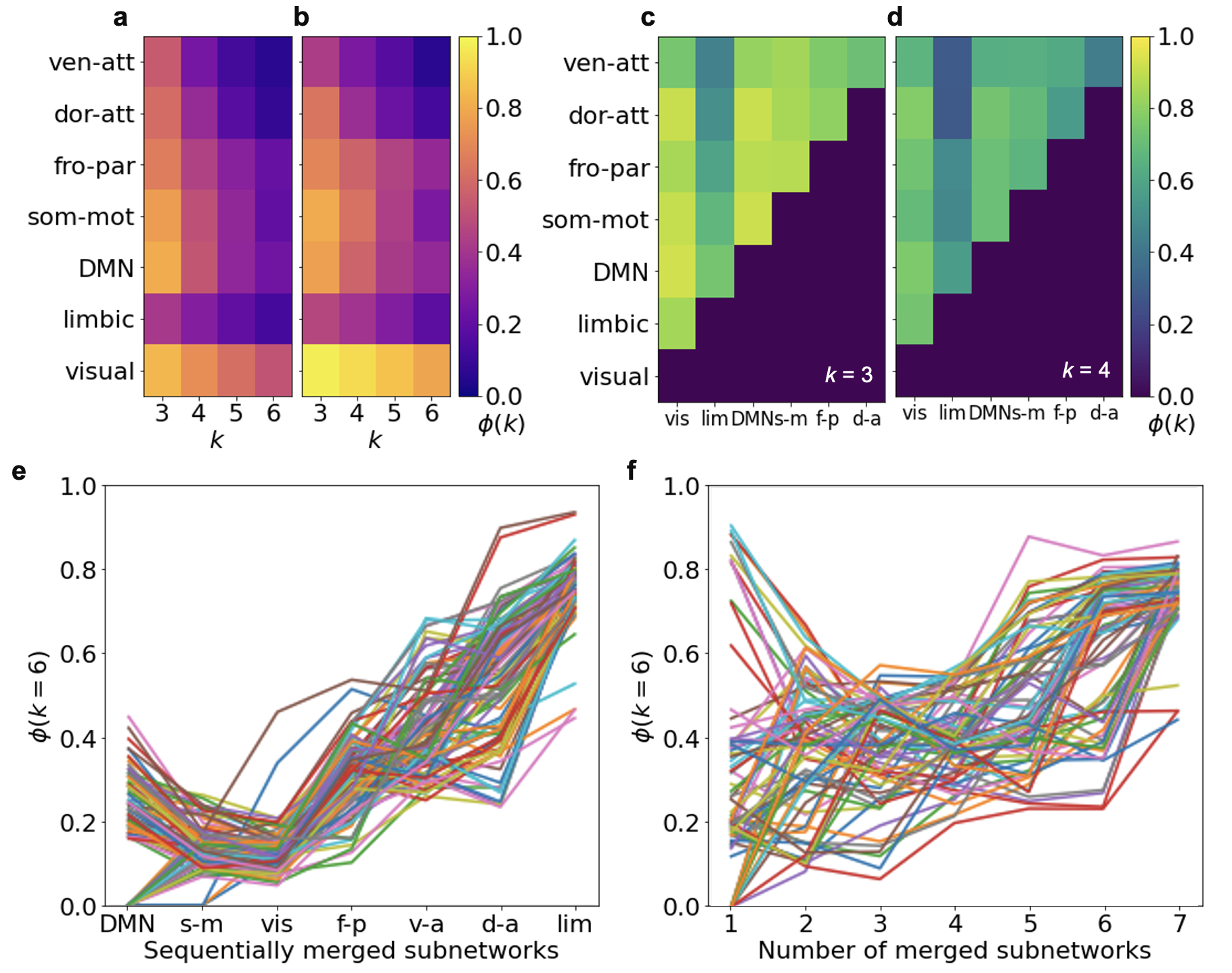}}
\caption{Analysis of $k$-clique clusters in functional subnetworks and their interactions. (a-b) Proportion of $k$-clique clusters in individual subnetworks: (a) considering only internal connections and (b) including both internal and external connections. (c-d) Proportion of $k$-clique clusters resulting from pairwise merging of subnetworks for clique orders (c) $k=3$ and (d) $k=4$. Values in (a-d) represent averages across $100$ networks. (e-f) Growth dynamics of $k$-clique cluster fraction $\phi(k)$ during sequential subnetwork merging: (e) ordered by subnetwork size and (f) in random order. Each curve represents a connectome.
}
\label{fig5}
\end{figure}

\paragraph{The emergence of high-order clique structures requires simultaneous interaction of several functional subnetworks.} 
Next, we investigated the clique structures in functional subnetworks. When examining each subnetwork in isolation, the maximum order of clique clusters with sufficient share does not exceed $k=3$-$4$, regardless of whether external connections are included (Fig.~\ref{fig5}a-b). Percolation of low-order $k$-cliques ($k=3$-$4$) is primarily a density-dependent feature, as discussed earlier. Thus, clique percolation in individual subnetworks is limited. The highest proportion of high-order clique structures is observed in the \emph{visual, default mode network (DMN), somato-motor}, and \emph{fronto-parietal} subnetworks (Fig.~\ref{fig5}a-b), which is directly related to the number of connections in these subnetworks.

When considering pairwise unions of subnetworks (including only internal connections for united pairs), the \emph{DMN} and \emph{somato-motor} networks demonstrate the most complementary connectivity with other subnetworks through clique structures (Fig.~\ref{fig5}c-d). For most combinations involving the \emph{DMN}, the resulting share of $k$-clique clusters exceeds the initial values for both the \emph{DMN} and the other subnetworks involved. 

With a sequential combination of all subnetworks, the total share of the $k$-clique cluster for high clique orders $k$ increases almost linearly (Fig.~\ref{fig5}e-f), regardless of the merging order (Fig.~\ref{fig5}f). At each step in (Fig.~\ref{fig5}e-f), $\phi(k)$ represents the proportion of vertices in the $k$-clique cluster relative to all vertices in the already combined subnetworks, not the entire connectome. Figure ~\ref{fig5}e shows $\phi(k)$ when subnets are connected sequentially in a fixed order (in descending order of the number of connections), while in Figure~\ref{fig5}f the subnetworks are connected in random order. A high-order clique cluster characteristic of human structural connectomes (e.g., $k=6$ of (Fig.~\ref{fig5}e-f)) only emerges when all subnetworks are connected. These findings show that the complementarity of external connections, i.e., the interaction of subnetworks, maintains the high-order clique percolation cluster. The complex, densely connected architecture observed in human connectomes through clique communities appears to be a distributed boundary of the interacting functional subnetworks rather than a property of individual ones. 

\subsection{Individual and common connections differ in their impact on structural connectivity.}

\begin{figure}[ht!]
\centerline{\includegraphics[width=1\linewidth]{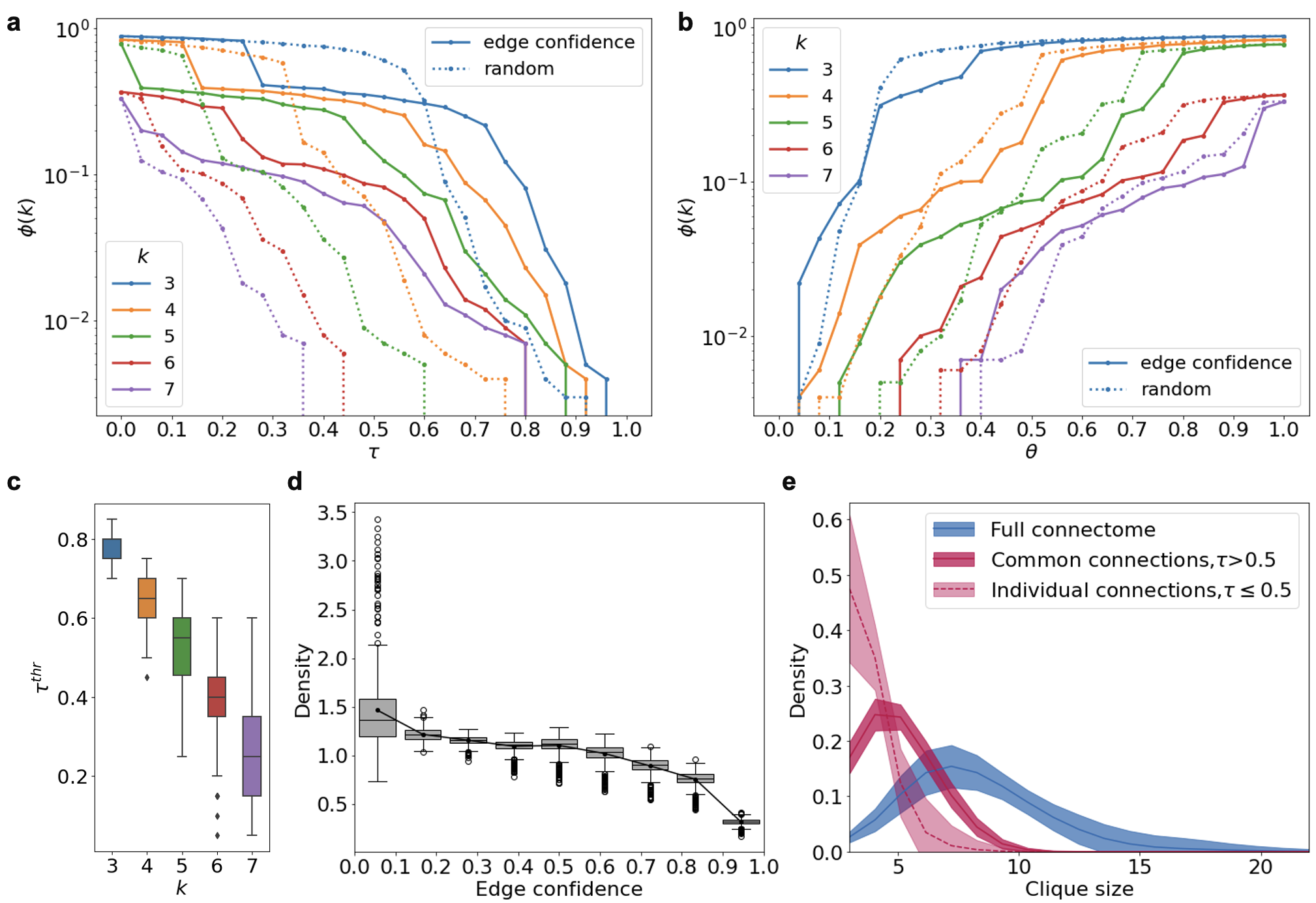}}
\caption{(a-b) The fraction of nodes ($\phi(k)$) included in $k$-clique clusters as a function of the cutoff threshold (a) $\tau$ for network decomposition and (b) threshold $\theta$ for inverse decomposition. (c) Distributions of critical destruction thresholds for clique clusters across structural connectomes, shown for different clique orders $k$. (d) Distribution of edge confidence values in individual human structural connectomes. (e) Distributions of maximal clique sizes in connectome networks.
}
\label{fig6}
\end{figure}

\paragraph{Edge confidence is associated with the sustainability of community structure connectivity in the human structural connectome.}
To understand the relation between edge confidence (connection weight showing frequency of occurrence in a connectome sample) and clique community formation, we analyzed the fraction of nodes ($\phi(k)$) included in $k$-clique clusters as a function of the cutoff threshold (Fig.~\ref{fig6}). We consider thresholds $\tau$ for network decomposition and thresholds $\theta$ for inverse decomposition (see the Methods section for details). 

The cutting occurs according to the edge confidence values, which are distributed in individual networks quite evenly, excepting the extremely small and large values (Fig.~\ref{fig6}d). The peak in the distribution of the lightest connections is most likely associated with errors in their determination, and they are also distinguished when studying the characteristic length of connections, constituting the overwhelming majority of the long-range domain (Fig.~\ref{fig1}b). 

Higher-order $k$-clique clusters ($k=5,6,7$) only percolate in the initial network state and rapidly disintegrate at small decomposition threshold values (Fig.~\ref{fig6}a). In contrast, percolation clusters for $k=3$ and $k=4$ encompass most of the network volume up to relatively high threshold values (Fig.~\ref{fig6}a, blue and orange solid lines). This behavior aligns with density-dependent $k$-clique percolation observed in random network models (see Methods, $k$-clique percolation section). The inter-hemispheric binding structure for $k=3$ clusters disintegrates at $\tau\approx 0.2$. 

For $k$-clique percolation clusters of small orders, we observe a narrow distribution of destruction threshold values (calculated from the peak of the derivative) across the entire sample of networks (Fig.~\ref{fig6}c). As the order $k$ increases, the spread of the destruction threshold increases (Fig.~\ref{fig6}c), indicating heterogeneity in the dense community structures within human connectomes.

Next, we conducted a comparative experiment replicating the decomposition procedure, removing edges randomly in the same proportion as in the original networks at each simulation step (Fig.~\ref{fig6}a, dotted lines). Random connection cutting resulted in faster decomposition of the $k$-clique cluster, with connectivity loss occurring even for small clique orders (Fig.~\ref{fig6}a, compare dotted and solid lines). In the inverse decomposition process, for small orders ($k=3$), most $k$-clique clusters emerge at low $\theta$-threshold values (Fig.~\ref{fig6}b). However, for higher orders ($k\geq 4$), growth in clique percolation cluster size only occurs at large $\theta$ values, corresponding to the addition of high-confidence connections. Interestingly, random edge addition turned out to be more effective in increasing clique cluster size during inverse decomposition, which can be seen by the cluster growth dynamics (Fig.~\ref{fig1}e, compare solid and dotted lines). Adding edges in a random order leads to earlier recovery of more common connections, further demonstrating their substantial contribution to clique cluster structure recovery. 

The observed difference between individual and common connections in importance for the clique structures connectivity is also confirmed by observation of the distributions of maximum cliques. In the initial network state, the characteristic clique size ranges within $5-10$ and reaches $20$ in the heavy right tail of the distribution (Fig.~\ref{fig6}e). We compared the distribution of maximal cliques if we cut half of the most individual or, on the contrary, half of the most common connections in connectomes. Comparison of the distribution shifts clearly shows the difference between the influence of individual and common connections on the characteristic clique sizes and confirms the special role of common connections for the connectivity of high-order clique structures (Fig.~\ref{fig6}e). Note that the effect is present against the fact that more connections remain in the network when cutting half with higher edge confidence values.

The observed patterns in clique size distribution and their dependence on edge weight suggest a hierarchical organization of brain connectivity, where stronger, more common connections help in maintaining large-scale network structure. This hierarchical organization may facilitate efficient information processing and integration across different brain regions \cite{park2013structural}.

\begin{figure}[ht!]
\centerline{\includegraphics[width=1\linewidth]{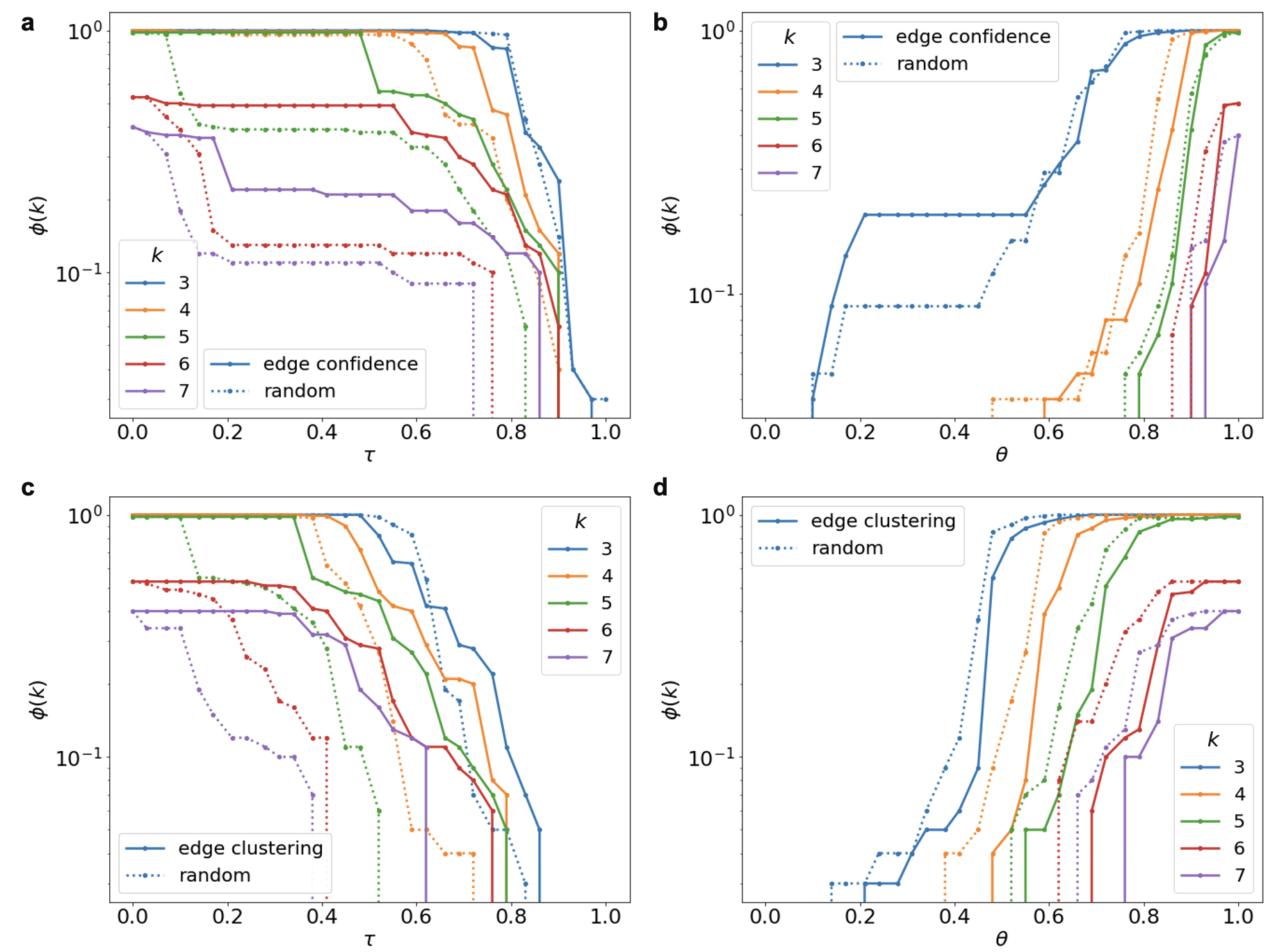}}
\caption{Model network decomposition (a, c) and inverse decomposition (d, b) using (a, b) generated sample-based edge confidence and (c, d) edge clustering coefficient weights. Solid lines represent the $k$-clique cluster size when edges are removed or added based on their weight; dotted lines show the cluster size when edges are randomly removed or added in the same proportion.
}
\label{fig7}
\end{figure}

We further calculated edge confidence for a sample of model networks. These networks were generated through the previously described process, increasing their transitivity to $0.4$ under fixed connection length constraints and with fixed node coordinates (the maximum connection length was set to $0.7$ in each of $200$ model networks with $100$ nodes and $1000$ edges). For both decomposition and inverse decomposition processes in model networks, edge confidence has a similar relation to the stability of clique cluster connectivity (Fig.~\ref{fig7}a-b). 

Network destruction also occurs later when compared to random edge removal in decomposition. During inverse decomposition, the cluster growth dynamics mirror those in connectomes; the earlier addition of heavy edges due to random selection leads to slide faster cluster volume recovery. Additionally, we explored another connection characteristic, the edge clustering coefficient, which represents the proportion of potential triangles associated with each edge, ranging from $0$ to $1$ \cite{valba2022} (Fig.~\ref{fig7}c-d). While the critical thresholds and the dynamics of network destruction and restoration differ due to variations in the weight distributions across network edges, the effect related to sustainability to decomposition is consistent for both model weights. It suggests that edge confidence and edge clustering coefficient serve as reliable indicators of a connection's importance in maintaining the integrity of clique communities. This relationship becomes pronounced in higher-order clique structures, showing the complex relationship between local network properties and global community organization.

\begin{figure}[ht!]
\centerline{\includegraphics[width=1\linewidth]{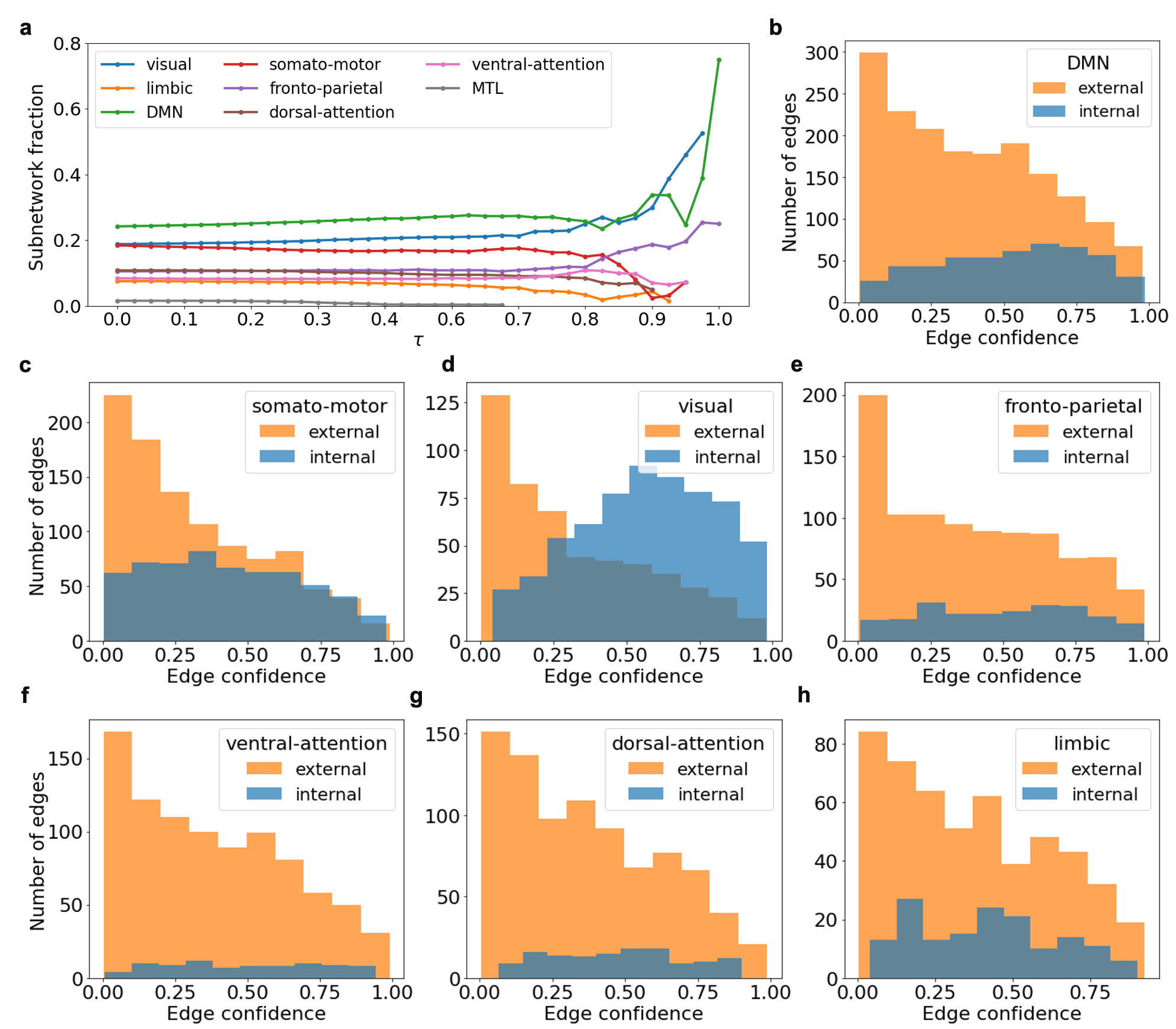}}
\caption{(a) The share made up of subnetwork nodes during decomposition by edge confidence value. An individual node is no longer taken into account if there are no connected edges of the subnetwork left. (b-h) Distributions of edge confidence for external and internal connections of subnetworks.
}
\label{fig8}
\end{figure}

During the decomposition process, subnetworks retain their relative share until the onset of critical collapse, when only the most common connections remain and the network loses connectivity (Fig.~\ref{fig8}a). The threshold for the onset of critical destruction of subnetworks is in the range at which $k=3$-$4$-clique percolation clusters disintegrate in individual connectomes and random graphs of equivalent density. This finding demonstrates the robustness of functional subnetworks up to a critical point.

\paragraph{Structural connections of interaction between functional subnetworks are more diverse and individual than internal ones.}

Functional subnetworks differ significantly not only in the number of external and internal connections but also in their characteristic edge confidence distributions (Fig.~\ref{fig8}b-h). The peak in the distribution of edge confidence in connectomes in the domain of the most individual connections turns out to be formed by external connections of subnetworks. Internal connections are more evenly distributed by the edge confidence parameter, showing a greater commonality in the internal structure of functional subnetworks in comparison with the individual diversity of their external connections. The \emph{visual} subnetwork stands out as the only one with more internal than external connections, with its internal connections showing a bias towards higher edge confidence values (Fig.~\ref{fig8}d). This may reflect the specialized and highly integrated nature of visual processing in the brain \cite{bullmore2009complex}. 

\section{Discussion}

\paragraph{The brain's structural connectome balances efficiency and constraints through high-order clique clusters, providing dense local connectivity in distributed brain networks.}

Using network science approaches, we demonstrated clique percolation of remarkably high order in human connectome networks. These structures reflect dense, highly overlapped structural communities distributed throughout the connectome. They also form distributed boundaries between interacting functional subnetworks, demonstrating a complex organization of brain connectivity. We suggest that these clusters emerge as a result of metric constraints on connection length, reflecting the brain's evolutionary need to balance connectivity efficiency with biological limitations. The observed order of the percolating clique cluster is anomalous for the characteristic connectome density and couldn't be explained as a density-dependent phenomenon. The resulting structural organization maintains a high density of local connectivity distributed across the connectome while preserving overall network sparsity. 

Our new network model, incorporating constraints on connection length, successfully replicated the observed structural properties. This model is based on the dynamical formation of the network structure through a process of random connection rewiring, increasing recurrent connections while maintaining network density. By growing recurrent connections, we restore the naturally observed triangle density, reflecting recurrent connectivity in connectomes. With certain restrictions on the connection length, we observe a structural phase transition, with the emergence of high-order $k$-cliques percolation. Such a dynamic is typically associated with critical phenomena and characterized by a continuous change in the order parameter (in this case, the size of the clique percolation cluster) as the control parameter is varied (connection length constraint). However, a more detailed analysis of this phenomenon is beyond the scope of the present research and requires further investigation to fully characterize its properties and implications for brain network organization. Without restrictions on connection length, the model demonstrates how the saturation of the network with triangles leads to excessive clustering, where dense structures occupy only a fraction of the network, reaching maximal density.

\paragraph{The observed structural clique clusters form a distributed structural boundary between interacting functional subnetworks.}

We examined the structural basis of functional brain subnetworks, which were identified with high separability and reproducibility in functional connectomes \cite{yeo2011organization}. We show that clique communities are not observed when considering functional subnetworks in isolation; they emerge only when subnetworks interact. The structural connections of subnetworks complement each other, resulting in a distributed clique structure. 

Functionally separable subnetworks are not well-defined structural clusters. In fact, the number of external connections (i.e., connections with other subnetworks) mostly for each subnetwork outnumbers its internal connections. Consequently, the structural organization of a functional subnetwork primarily consists of external connections between subnetworks, while internal connections do not form dense structures. In essence, the observed clique clusters form a boundary between interacting functional subnetworks, demonstrating the complementary nature of structural connections among these interacting subnetworks. Such structural boundaries, distributed throughout the connectome,  underline the complex interplay between functional and structural networks. Such internal organization could help integrate information across different brain regions and show the hierarchical structure of brain networks. It suggests a model of brain function where specialized processing within subnetworks is balanced with global integration across the entire connectome.

It was proposed that high-order cliques in structural brain networks could serve as building blocks for cognitive processes, potentially supporting information integration across different brain regions \cite{sizemore2019importance}. The existence of these structures might contribute to the brain's resilience and adaptability \cite{bullmore2012economy}. This finding aligns with recent research suggesting that brain networks exhibit small-world properties and rich-club organization, facilitating efficient information processing and integration \cite{watts1998collective, van2011rich}.


\paragraph{The frequency of structural connection occurrence in a population is associated with phenomena in individual brain networks.}

By analyzing the difference between individual and general connections for connectomes from the sample, we observed the distinguished role of the latter in the formation and connectivity of clique structures. The edge clustering coefficient, which is another connection characteristic, had a similar property, which reveals a straightforward effect: connections that contribute most significantly to local triangle density have the greatest impact on forming large cliques and maintaining connectivity within the clique communities.

Another noteworthy finding is the difference in edge confidence distributions between internal and external connections of subnetworks. External connections lack the characteristic growth in distribution as edge confidence weakens; instead, their distribution is nearly uniform. This suggests that internal connections are more repetitive and common across the connectome sample, regardless of their numerical inferiority to external connections. Considering that edge confidence reflects the likelihood of a particular structural connection occurring in a set of connectome networks (i.e., its prevalence in a population), we found that statistically frequent structural connections contribute most to the connectivity of high-order clique structures. Thus, connection properties correlate with their frequency of occurrence in the connectome set, even though these connections in individual connectomes form throughout participants' lifetimes. We observe a surprising commonality of the basic structural organization, possibly determining the similarity of our functional organization, and at the same time a great diversity and variability of connections that complement the basic structures. This phenomenon bears a resemblance to the concept of ergodicity in physics, where observing a system's property over time (time averaging) is equivalent to observing a sample (ensemble averaging) \cite{kastner2008phase}.

\section{Conclusion}

The $k$-clique percolation approach provides a powerful tool for unraveling relationships between structural and functional brain networks. The discovery of high-order $k$-cliques organization in the human structural connectome expands our understanding of brain connectivity. These complex structures support the hypothesis that the brain is not merely a collection of isolated functional modules but rather an intricately connected system where densely interlinked regions support cognitive processes across multiple domains. This finding demonstrates that higher-order clique communities maintain both local and global network integration, potentially supporting critical cognitive functions. 

\section{Acknowledgements}
This work/article is an output of a research project implemented as part of the Basic Research Program  at the National Research University Higher School of Economics (HSE University). This research was supported in part through computational resources of HPC facilities at HSE University. This work has benefited from a French government grant managed by the Agence Nationale de la Recherche under the France 2030 program, reference ANR-23-IAHU-0003.

\printbibliography
\end{document}